\documentclass[12pt]{article}
\usepackage{amsmath}
\usepackage{amssymb}
\usepackage{amstext}
\usepackage{amscd}
\usepackage{amsfonts}
\usepackage{graphicx}
\usepackage{color}
\begin{document}
\newcommand{\nd}{\noindent}
\newcommand{\nl}{\newline}
\newcommand{\be}{\begin{equation}}
\newcommand{\ee}{\end{equation}}
\newcommand{\ben}{\begin{eqnarray}}
\newcommand{\een}{\end{eqnarray}}
\newcommand{\nn}{\nonumber \\}
\newcommand{\ii}{\'{\i}}
\newcommand{\pp}{\prime}
\newcommand{\expq}{e_q}
\newcommand{\lnq}{\ln_q}
\newcommand{\quno}{q-1}
\newcommand{\qunoinv}{\frac{1}{q-1}}
\newcommand{\tr}{{\mathrm{Tr}}}
\newcommand{\1}{\beta}
\newcommand{\2}{\gamma}
\newcommand{\3}{\rho}

\title{{\bf 3D EFFECTS OF THE ENTROPIC FORCE}}

\author{{\small A. Plastino$^1$, M. C. Rocca$^1$, and  G. L. Ferri $^2$} \\
\small{$^1$ La Plata National University and
  Argentina's National Research Council}\\
\small{(IFLP-CCT-CONICET)-C. C. 727, 1900 La Plata - Argentina}\\
$^2$ \small{Faculty of Exact and Natural Sciences,  La Pampa
National University} \\{\small Uruguay 151, (6300) Santa Rosa, La
Pampa, Argentina} }

\date{\today}

\maketitle

\begin{abstract}
\nd
     This work analyzes the classical statistical mechanics associated to  phase-space curves in three dimensions.
     Special attention is paid to the entropic  force. Strange effects like  confinement, hard core, and asymptotic
      freedom are uncovered.
      Negative specific heats, that were previously seen to emerge in a one-dimensional setting, disappear in 3D,
      and with them, gravitational effects of the entropic force.

\vskip 3mm

\nd {\small {\bf Keywords:} Phase-space curves, Entropic force,
Confinement, Hard core, Asymptotic freedom}.

\end{abstract}

\newpage

\normalcolor
\section{Introduction}

\nd  A previous {\it one-dimensional} study concerning the
classical statistical mechanics of arbitrary phase-space curves
$\Gamma$ \cite{uno1} was shown to uncover some strange (for a
classical setting), like confinement and asymptotic freedom. Of
course, by confinement one alludes to the physical phenomenon that
impedes isolation of color charged particles (such as quarks),
that cannot be isolated singularly, while asymptotic freedom is a
property of some gauge theories that originates bonds between
particles to become asymptotically weaker as distance decreases.
\vskip 3mm

\nd The classical analysis of \cite{uno1}   provided a simple
entropic {\bf mechanism} for these ``strange" phenomena. Remind
the reader that the entropic force is a {\it phenomenological} one
arising from some systems' statistical tendency to increase their
entropy \cite{pol1,pol2,verlinde,path,dewar}, without  appealing
to any specific underlying microscopic interaction.  The most
important example is the elasticity of a freely-jointed polymer
molecule
  (see, for examples, \cite{pol1,pol2} and references therein).
Issues revolving this force were made popular recently by
Verlinde, who argued that gravity can also be thought of as an
entropic force \cite{verlinde}. The Coulomb force enters such game
too  \cite{wang}, etc. Note that there exists an exact solution
for the static force between two black holes at the turning points
in their binary motion \cite{uno} and also that research
concerning the entanglement entropy of two black holes associates
 an  entanglement entropic force \cite{dos}. A causal
path entropy (causal entropic forces) has been recently made
responsible for links between intelligence and entropy
\cite{path}. \vskip 2mm

\nd Here we revisit \cite{uno1} and try to extend its results to a
more ``real" three-dimensional setting by appeal to a simple model
(quadratic Hamiltonian in phase-space). We will  show that
confinement also arises here from entropic forces. \vskip 2mm

\nd Of course, quadratic Hamiltonians are customarily appealed to
 both in classical and in quantum mechanics. For them, the correspondence between classical and
quantum mechanics is exact. However,  explicit formulas are not
necessarily trivial. A good knowledge of quadratic Hamiltonians is
of utility  in dealing with more general quantum Hamiltonians for
the semiclassical approach. These Hamiltonians are also important
in partial differential equations, because they give non trivial
examples of wave propagation phenomena. Finally, they help to
understand properties of more complicated Hamiltonians used in
quantum theory.

\nd We will deal with  quadratic Hamiltonians in a classical
setting so as to  to discern  whether interesting features that
emerge in one dimension appear also in 3D when studying  the
entropic force along phase-space curves.

\section{Formalism}

\nd We recapitulate here the formalism expounded in \cite{uno1}
and  consider a typical, $n$-dimensional harmonic oscillator-like
Hamiltonian  in thermal contact with a heat-bath at the inverse
temperature $\beta$ (that will be kept constant throughout).

 \be
\label{ep1.1}    H(p,q)= p^2+ q^2, \ee
\[p^2=p_1^2+p_2^2+\cdot\cdot\cdot p_n^2\;\;\;
q^2=q_1^2+q_2^2+\cdot\cdot\cdot q_n^2\] \nd where $p$ and $q$ have
the same dimensions (natural units, those of $H$, obviously. We
wish to avoid dealing with a tensor $g_{ij}$). \nd The
corresponding partition function is given by
\cite{patria,lavenda,katz}
\[Z(\beta)=\int\limits_{-\infty}^{\infty}\,e^{-\beta
H(p,q)}\;d^npd^nq=\] \be \label{ep1.2} \left[\frac {2\pi^{\frac
{n} {2}}} {\Gamma(n/2)}\right]^2
\int\limits_0^{\infty}\int\limits_0^{\infty} e^{-\beta
H(p,q)}p^{n-1}q^{n-1}\;dp\;dq. \ee One has integrated over the
angles in phase space. Employ the fact that the total microscopic
energy is \be \label{ep1.3} U=p^2+q^2, \ee and  make the change of
variable $p=\sqrt {U-q^2}$. An important  result
follows

\begin{equation} \label{es1.1} \frac {2\pi^n}
{\Gamma^2(n/2)}\int\limits_0^{\infty}
\int\limits_0^{\sqrt{U}}e^{-\beta U} p^{n-1} (U-p^2)^{\frac {n-2}
{2}}\;dp\;dU=
\end{equation}
\begin{equation}
\label{es1.2} \frac {2\pi^n} {\Gamma^2(n/2)}\int\limits_0^{\infty}
\int\limits_0^{\sqrt{U}}e^{-\beta U} x^{\frac {n-2} {2}}
(U-x)^{\frac {n-2} {2}}\;dx\;dU,
\end{equation}
where we  changed  variables in the fashion  $p^2=x$. Computing
(\ref{es1.2}) in  $x$ using \cite{gra1} gives a basic result
\begin{equation}
\label{es1.3} Z(\beta)=\frac {\pi^n} {\Gamma(n)}
\int\limits_0^{\infty} U^{n-1} e^{-\beta U}\;dU.
\end{equation}
Evaluating (\ref{es1.3}) in $U$ with the help of \cite{gra2}
produces
\begin{equation}
\label{es1.4} Z(\beta)=\frac {\pi^n} {{\beta}^n}.
\end{equation}
Similarly, we obtain  for the mean value of the energy
\[<U(p,q)>(\beta)=\frac {1} {Z(\beta)}
\int\limits_{-\infty}^{\infty}H(p,q)e^{-\beta H(p,q)}\;d^npd^nq=\]
\be \label{ep1.4} \frac {\pi^n} {\Gamma(n) Z(\beta)}
\int\limits_0^{\infty} U^n e^{-\beta U}\;dU=\frac {n} {\beta}, \ee
and for the entropy $S$
\[S(\beta)=\frac {1} {Z(\beta)} \int\limits_{-\infty}^{\infty}
[\ln Z(\beta) + \beta H(p,q)]e^{-\beta H(p,q)}\;dpdq=\] \be
\label{ep1.5} \frac {\pi^n} {\Gamma(n)Z(\beta)}
\int\limits_0^{\infty} \{\ln [Z(\beta)]U^{n-1} e^{-\beta
U}\;dU+<U>=\ln [Z(\beta)] + n. \ee Importantly enough,  the
integrands (\ref{es1.3}),(\ref{ep1.4}), and (\ref{ep1.5}) {\it are
exact differentials}.

\subsection{Entropy along a path $\Gamma$}

 \nd  This is our central concept. We recapitulate here the associated
main ideas, advanced in \cite{uno1}. Of course, one assumes in
contact with a reservoir an the fixed inverse temperature $\beta$.
Path entropies (phase space curves) have been considered in, for
example, Ref. \cite{uno1,path,dewar}. We deal in this work with a
{\it related, but not identical concept}. Following ideas of
\cite{uno1}, we consider  a particle moving in phase space,
centering attention on its entropy
 computed as it traverses some phase space path $\Gamma$ that starts at the origin and ends at some  arbitrary point
($p_1(q^0_1),...,p_n(q^0_1),q_2(q_1^0),...,q_n(q_1^0))$.
  $\Gamma$ is thus parameterized by the phase-space variable $q_1$. Our goal is to define the thermodynamic variables  along these
phase-space curves $\Gamma$.  All our calculations  are of a
microscopic character. No macrostates are used.  Generalizing the
exact differentials-integrands r (\ref{es1.3}),(\ref{ep1.4}), and
(\ref{ep1.5}) to curves $\Gamma$, we introduce

\begin{itemize}

\item  The partition function as a function of $\beta$ and of a
curve $\Gamma$

\be \label{ep2.1} Z(\beta,\Gamma)=\frac {\pi^n} {\Gamma(n)}
\int\limits_{\Gamma} U^{n-1}(p,q)
e^{-\beta U(p,q)}\;dU(p,q). \ee

\item  The mean energy as

\be \label{ep2.2} <U(p,q)>(\beta,\Gamma)=\frac {\pi^n}
{\Gamma(n)Z(\beta,\Gamma)} \int\limits_{\Gamma} U^n(p,q) e^{-\beta
U(p,q)}\;dU(p,q). \ee

\item   Our path  entropy is then defined

\be \label{ep2.3} S(\beta,\Gamma)=\frac {\pi^n}
{\Gamma(n)Z(\beta,\Gamma)}\int\limits_{\Gamma} \{\ln [Z(\beta,\Gamma)]
+\beta U(p,q)\}U^{n-1}(p,q) e^{-\beta U(p,q)}\;dU(p,q).   \ee

\end{itemize}

\nd   As in \cite{uno1}, we consider curves, parameterized as a
function of the independent variable $q_1$, passing through the
origin, for which we have $p(0)=0$ and $q(0)=0$, and, thus,  for
any temperature $T$, $U(0,0)=0$.  This picture can be always
arrived at after an adequate coordinates-change \cite{uno1}. If we
take into account that a) the integrands are exact differentials
and b) the integrals are independent of the curve's shape and only
depend on their end-points $q^0_1$, we deduce

\nd i) For the partition function
\[Z(\beta,q_1^0)=\frac {\pi^n} {\Gamma(n)} \int\limits_0^{q_1^0}
U^{n-1}[p(q_1),q(q_1)] e^{-\beta
U[p(q_1),q(q_1)]}\;dU[p(q_1),q(q_1)]\] and evaluating the integral
with the help of \cite{gra3} \be \label{ep2.4}
Z(\beta,q_1^0)=\left\{\frac {\pi^n} {\beta^n}-e^{-\beta
U[p(q_1^0),q(q_1^0)]} \sum\limits_{s=0}^{n-1}\frac {\pi^n} {s!}
\frac {U^s[p(q_1^0),q(q_1^0]} {\beta^{n-s}}\right\}. \ee ii) For
the mean value of the energy \be \label{ep2.5}
<U(p,q)>(\beta,q_1^0)=\frac {\pi^n}
{\Gamma(n)Z(\beta,q_1^0)}\int\limits_0^{q_1^0} U^n[p(q_1),q(q_1)]
e^{-\beta U[p(q_1),q(q_1)]}\;dU[p(q_1),q(q_1)], \ee which gives
\begin{equation}
\label{ep2.6} <U(p,q)>(\beta,q_1^0)=\frac {\pi^n}
{Z(\beta,q_1^0)}\left\{ \frac {n} {\beta^{n+1}}-e^{-\beta
U[p(q_1^0),q(q_1^0)]} \sum\limits_{s=0}^n \frac {n} {s!} \frac
{U^s[p(q_1^0),q(q_1^0)]} {\beta^{n-s+1}}\right\}
\end{equation}
iii) For the entropy
\[S=\frac {\pi^n}{\Gamma(n)Z(\beta,q_0)}
\int\limits_0^{q_1^0} \{\ln Z(\beta,q_1^0) +U[p(q_1),q(q_1)]\}
U^{n-1}[p(q-1),q(q_1)]\times\]
\be
\label{ep2.7}
e^{-\beta U[p(q_1),q(q_1)]}\;dU[p(q_1),q(q_1)],
\ee
whose result is
\[S(\beta,q_1^0)=\frac {\pi^n} {Z(\beta,q_1^0)}
\left\{\ln [Z(\beta,q_1^0)]\left[\frac {1} {\beta^n}- e^{-\beta
U[p(q_1^0),q(q_1^0)]} \sum\limits_{s=0}^n\frac {1} {s!} \frac
{U^s[p(q_1^0),q(q_1^0)]} {\beta^{n-s}} \right]+\right.\] \be
\label{ep2.8} \left.\frac {n} {\beta^n}- e^{-\beta
U[p(q_1^0),q(q_1^0)]} \sum\limits_{s=0}^n\frac {n} {s!} \frac
{U^s[p(q_1^0),q(q_1^0)]} {\beta^{n-s}}\right\}. \ee

\nd Whenever $q_1^0\rightarrow \infty$ (\ref{ep2.4}),
(\ref{ep2.6}), and (\ref{ep2.8}) reduce to (\ref{es1.4}),
(\ref{ep1.4}), and (\ref{ep1.5}), respectively. Note again that
the integrands in (\ref{ep2.4}), (\ref{ep2.6}), and (\ref{ep2.8})
are {\it exact differentials}. We insist on the fact that  1)
these integrals become independent of the path $\Gamma$  (i.e.,
the same for any $\Gamma$), and 2) if one redefines the
coordinate-system  in such a way that the starting point of
$\Gamma$ coincides with the origin, their values {\bf will depend
only on the end-point of the path}. Thus, they are functions of
the {\bf microscopic} state (at least for the HO-Hamiltonian).
   We can call the entropy and the mean energy evaluated above as
 {\it microscopic thermodynamic potentials} (for the HO).
  Remind that we are in contact with a reservoir an
the fixed inverse temperature $\beta$.  The simplest possible
path-forms are straight lines connecting the origin with
($p(q_1^0), q(q_1^0)$).

\nd It was shown in \cite{uno1}, that our thermodynamics along
phase-space curves does make physical sense because it accounts
for an equipartition theorem. It was there  encountered that

\be <q_i^2>=<p_i^2>=\frac {<U>} {2n}=\frac {1} {2\beta}, \ee that is,
classical equipartition.

\nd It was also demonstrated in \cite{uno1} that adiabatic paths
exist, where an adiabatic path is, of course,  one such that $S=$
constant along it. From (\ref{ep2.8}) we obtain the condition $S=$
constant translates into, for example \be \label{ep3.2}
\beta=C_1\;\;\; U[p(q_1^0),q(q_1^0)]=C_2, \,\,\,{\rm independently
\,\,\,of} \,\,\,q_1^0. \ee $C_1=\beta$ is constant by the very
reservoir's notion.

\section{Our main interest: the entropic force}

\nd This is our main topic. According to (\ref{ep2.7}), the
entropic force is given by Eq. (3.3) of \cite{verlinde} that reads
$F_e dx= TdS $. In our case this translates as \cite{uno1}  \be
\label{ep4.1bis } F_{ei} dq_i=\frac {1} {\beta}\frac {\partial S}
{\partial q_i} dq_i, \ee \be \label{ep4.1} F_{ei}=\frac {1} {\beta
Z}\frac {\partial Z} {\partial q_i}+ \frac {\partial <U>}
{\partial q_i} \ee where:
\[\frac {\partial <U>} {\partial q_i}=
-\frac {<U>} {Z} \frac {\partial Z} {\partial q_i}+
\frac {\partial U[p(q_1^0),q(q_1^0)]} {\partial q_i}
e^{-\beta U[p(q_1^0),q(q_1^0)]}\times\]
\begin{equation}
\label{ep4.2}
\frac {\pi^n} {Z}\left[\sum\limits_{k=0}^n \frac {n} {k!}
\frac {U^k[p(q_1^0),q(q_1^0)]} {\beta^{n-k}}-
\sum\limits_{k=1}^n \frac {n} {(k-1)!}
\frac {U^{k-1}[p(q_1^0),q(q_1^0)]} {\beta^{n-k+1}}\right]
\end{equation}
and with:
\[\frac {\partial Z} {\partial q_i}=\pi^n
\frac {\partial U[p(q_1^0),q(q_1^0)]} {\partial q_i}
 e^{-\beta U[p(q_1^0),q(q_1^0)]}\times\]
\begin{equation}
\label{ep4.3}
\left[\sum\limits_{k=0}^{n-1} \frac {1} {k!}
\frac {U^k[p(q_1^0),q(q_1^0)]} {\beta^{n-k-1}}-
\sum\limits_{k=1}^{n-1} \frac {1} {(k-1)!}
\frac {U^{k-1}[p(q_1^0),q(q_1^0)]} {\beta^{n-k}}\right]
\end{equation}
The trajectory's end-point is free to move in phase-space. More
generally, for $U=p^2+q^2$ one has again
\[F_{ei}=\frac {1} {\beta Z}\frac {\partial Z} {\partial q_i}+
\frac {\partial <U>} {\partial q_i}\]
where:
\[\frac {\partial <U>} {\partial q_i}=
-\frac {<U>} {Z} \frac {\partial Z} {\partial q_i}+
2q_i
e^{-\beta (p^2+q^2)}\times\]
\begin{equation}
\label{ep4.4}
\frac {\pi^n} {Z}\left[\sum\limits_{k=0}^n \frac {n} {k!}
\frac {(p^2+q^2)^k} {\beta^{n-k}}-
\sum\limits_{k=1}^n \frac {n} {(k-1)!}
\frac {(p^2+q^2)^{k-1}} {\beta^{n-k+1}}\right]
\end{equation}
and with
\begin{equation}
\label{ep4.5} \frac {\partial Z} {\partial q_i}=2\pi^nq_ie^{-\beta
(p^2+q^2)} \left[\sum\limits_{k=0}^{n-1} \frac {1} {k!} \frac
{(p^2+q^2)^k} {\beta^{n-k-1}}- \sum\limits_{k=1}^{n-1} \frac {1}
{(k-1)!} \frac {(p^2+q^2)^{k-1}} {\beta^{n-k}}\right].
\end{equation}

\section{The three-dimensional scenario}

\nd In \cite{uno1} only the one-dimensional case was considered in
detail. In particular, we emphasize that we treat the entropic
forces in
      three-dimensional fashion. This is to be compared to the pioneer effort by Verlinde
      \cite{verlinde}, where the pertinent treatment is one-dimensional. We ask now, and this is our leit-motiv here, can the
findings in \cite{uno1} be extended to 3D? The answer is of a
mixed nature. \vskip 3mm
 \nd  We will present as evidence 3-dimensional plots  for three temperature
regimes, namely,
\begin{itemize}

\item Low temperatures, $\beta=4$ (Figs. 1-2),

\item Intermediate temperatures, $\beta=1$ (Figs. 3-4),

\item High temperatures, $\beta=0.2$  (Figs. 5-6).

\end{itemize}

 \nd  We plot the
 entropic force $F_e$ for, respectively, $\beta = 0.2, 1.0,$
and $4.0$ in Figs. 1, 2, and 3, respectively.  In these plots,
$q=\sqrt{[q_x^2+q_y^2+q_z^2]}$ and $p=\sqrt{[p_x^2+p_y^2+p_z^2]}$.
Each surface is parameterized by a fixed value of, respectively,
$q_x= 0.01, 0.05, \,\, 0.10,\,\,  0.20,\,\, 0.30,\,\, 0.40,\,\,
0.50$ top-wards.

\nd We see that there is an infinitely repulsive barrier (hard
core) near (but not {\bf at}) the origin. In the immediate
vicinity of the origin it can analytically be shown that the force
vanishes. It also tends to zero at long distances from the
hard-core. The conjunction between these facts, as in the one
dimensional instance, yields both confinement and asymptotic
freedom via a simple classical mechanism.


\nd Of course, our particle not only feels the $F_e-$influence but
also that of the negative gradient of the HO potential . Thus, it
is affected by a total force $F_{T}= F_e+F_{HO}$. The pertinent
expression is \cite{uno1}

\be F_T=\frac {1} {\beta Z}\frac {\partial Z} {\partial q_i}+
\frac {1} {2}
\frac {\partial <U>} {\partial q_i}
\ee
where
\be
F_{HO}=-\frac {1} {2}
\frac {\partial <U>} {\partial q_i}
\ee

 \nd  We plot this total force for, respectively, $\beta = 0.2, 1.0,$
and $4.0$ in Figs. 4, 5,  and 6,  respectively. It is seen that
the essential features described in the first three plots do not
suffer any appreciable qualitative change.

\vskip 3mm

\nd  The {\it specific heat},     that is, the derivative of the
mean energy with respect to the temperature at constant volume, is
easily seen, as shown in \cite{uno1} to be

\[ C=-\frac {1} {Z}\frac {\partial Z} {\partial T} <U>+
\frac {k\pi^n} {Z} \left[n(n+1)(kT)^n-e^{\frac {p^2+q^2} {kT}}
\sum\limits_{s=0}^n\frac {n} {s!}(p^2+q^2)^{s+1}
(kT)^{n-s-1}
\right.\]
\be
\left.
-e^{\frac {p^2+q^2} {kT}}
\sum\limits_{s=0}^n\frac {n(n-s+1)} {s!}(p^2+q^2)^s
(kT)^{n-s}\right]
\ee
 independently of the curve $\Gamma$, with
$k=k_{Boltzmann}$ and
\[\frac {\partial Z} {\partial T}=
k\pi^n \left[n(kT)^{n-1}-e^{\frac {p^2+q^2} {kT}}
\sum\limits_{s=0}^{n-1}\frac {(p^2+q^2)^{s+1}} {s!}
(kT)^{n-s-2}
\right.\]
\be
\left.
-e^{\frac {p^2+q^2} {kT}}
\sum\limits_{s=0}^{n-1}\frac {n-s} {s!}(p^2+q^2)^s
(kT)^{n-s-1}\right]
\ee

 \nd  Figs. 7 depicts $C$ for, respectively, $\beta= 0.2$, $1$, and $2$. Fig. 8 displays $C$ in an extended $(q,p)$-range,
 for $\beta=0.2$, in order to better appreciate the classical limit. Fig 9 is identical to Fig. 8, but for 2 and 4 dimensions. The
specific heat does not  changes sign, being always positive. It
tends to $C=3k$ (k being Boltzmann's constant) at infinity, the
``classical result". Negative
 specific heats are perhaps the most distinctive  thermodynamic feature of
 self-gravitating systems \cite{binney}. Here, our entropic discourse
 does not back  Verlinde's ideas, that are however based on a one-dimensional picture
 \cite{verlinde}.Remark that the gravitational potential of a
 point mass located at the origin (solution of the Laplace equation) depends on the dimension. It is
 a constant for N=1, a logarithm for N=2, and a power law for
 $N>2$.

\section{Discussion}

\nd We considered a particle attached to the origin by a spring
and discussed entropic-force effects. Although we focused
attention upon arbitrary phase space curves $\Gamma$, most of our
effects were independent of the specific path $\Gamma$.
 Our statistical mechanics-along-curves concept is seen to make sense because
 the equipartition theorem is valid for it. Most of our
 one-dimensional discourse, expounded in \cite{uno1}, remains
 valid in 3D, with  an important exception. We do not find negative specific
 heats. contrary what happens in one dimension. Thus, no link to
 gravitation can be established. The specific heats vanish ar zero
 temperature, of course, as can be analytically determined.

 \nd  From Figs. 1-6 we gather the
entropic force diverges at short distances from the origin
(hard-core effect), but vanishes both just there
 and at infinity, so that, with some abuse of language one may speak of ``asymptotic freedom".
\nd The entropic force is repulsive.  As stated above, at long
distances  from the origin the entropic force tends to vanish.

\vskip 3mm

\nd Entropic confinement is the most remarkable effect that our
classical entropic force-model exhibits. Independently  of whether
our model is realistic or not, it does provide a classical
confinement mechanism. The present considerations should encourage
non-classical explorations regarding the entropic force. \vskip
3mm

\nd Finally, when we couple the entropic force effects with those
of the HO-potential we are not  able to discern significant new
features. We have presented here somewhat counter-intuitive
results.

\nd Summing up, the specific heat plots clearly exhibit what may
be regarded as the most significant effect of the entropic force.
In the vicinity of the particles's location (the origin), it
depresses the $C-$value from its classical constant value, forcing
it to vanish at the origin. Remarkably enough, one easily
ascertains, in analytic fashion, that $C$ adopts its classical
constant value everywhere at $T=0$, implying that, at that
temperature, the entropic force vanishes.

\nd Further work should try to incorporate the interesting
entropic notions developed by Sadhukhan and Bhattacharjee in
\cite{SMB}.

\newpage

\begin{figure} [htbp]
\centerline{
{\scalebox{0.7}{\includegraphics{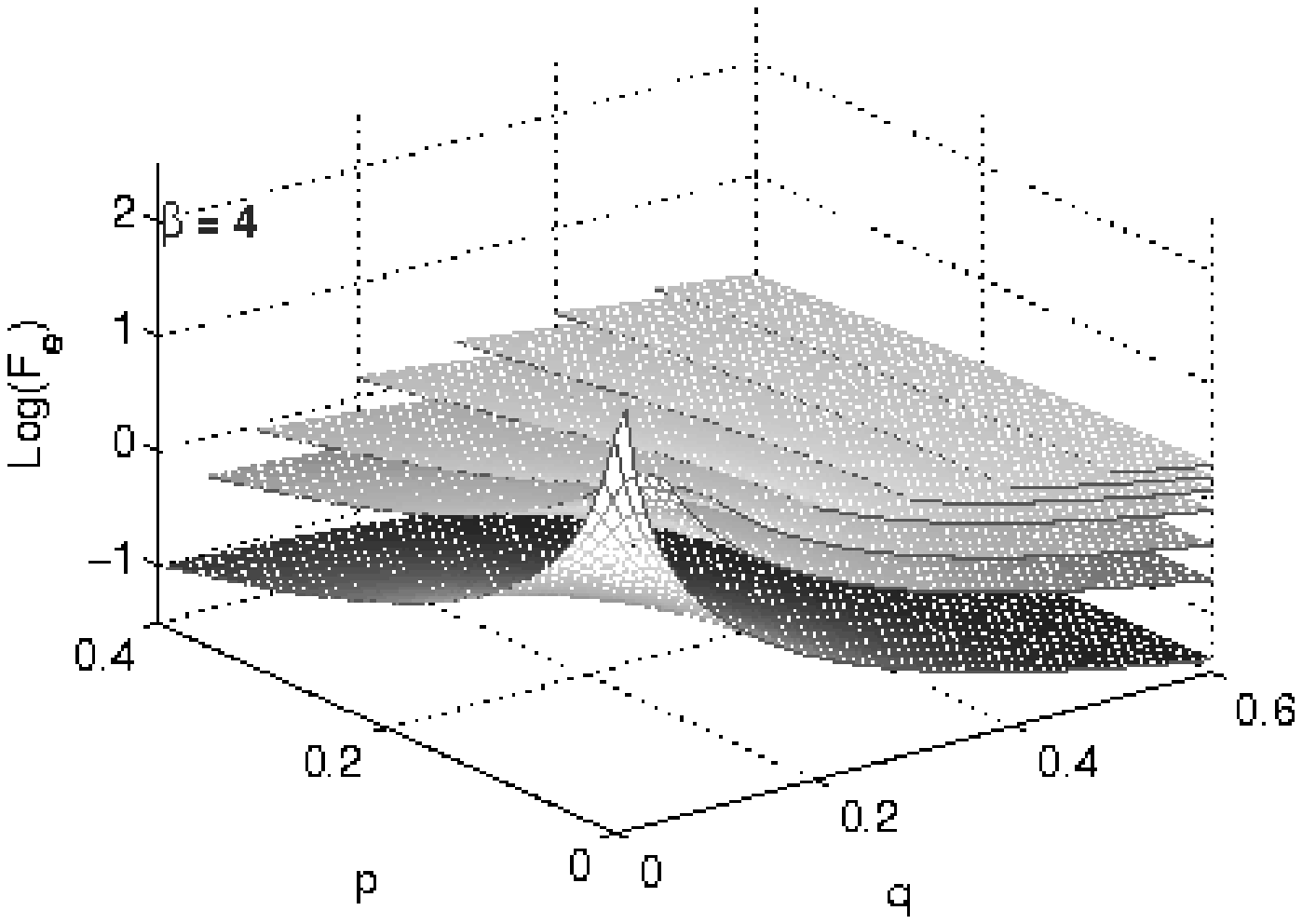}}} }
\caption{$\log_{10} F_e$ for $\beta=4$.}. For further details, see
text. \label{FigEntro40}
\end{figure}

\newpage

\begin{figure} [htbp]
\centerline{
{\scalebox{0.7}{\includegraphics{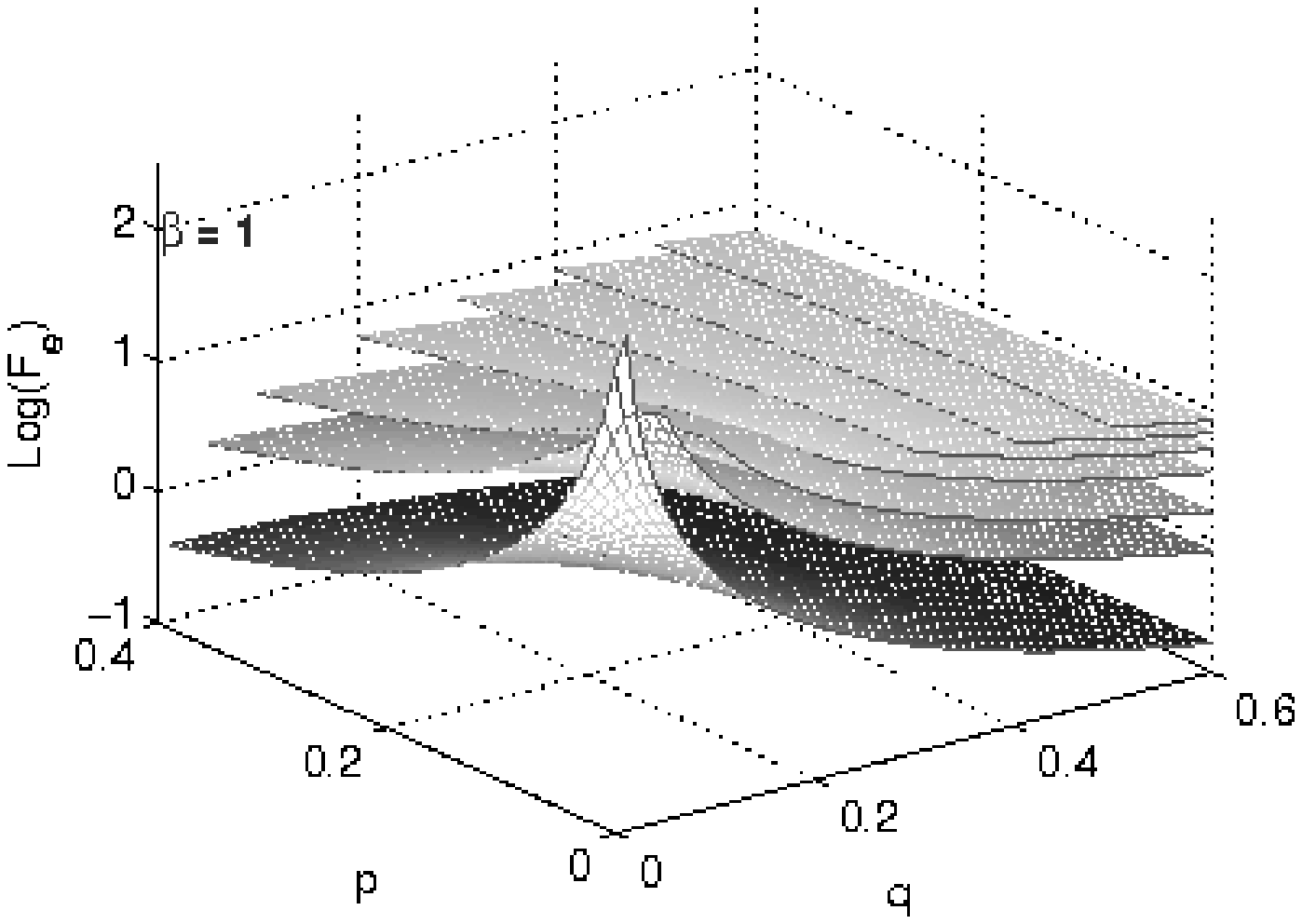}}} }
\caption{$\log_{10} F_e$ for $\beta=1$.} For further details, see
text.  \label{FigEntro10}
\end{figure}

\newpage

\begin{figure} [htbp]
\centerline{
{\scalebox{0.7}{\includegraphics{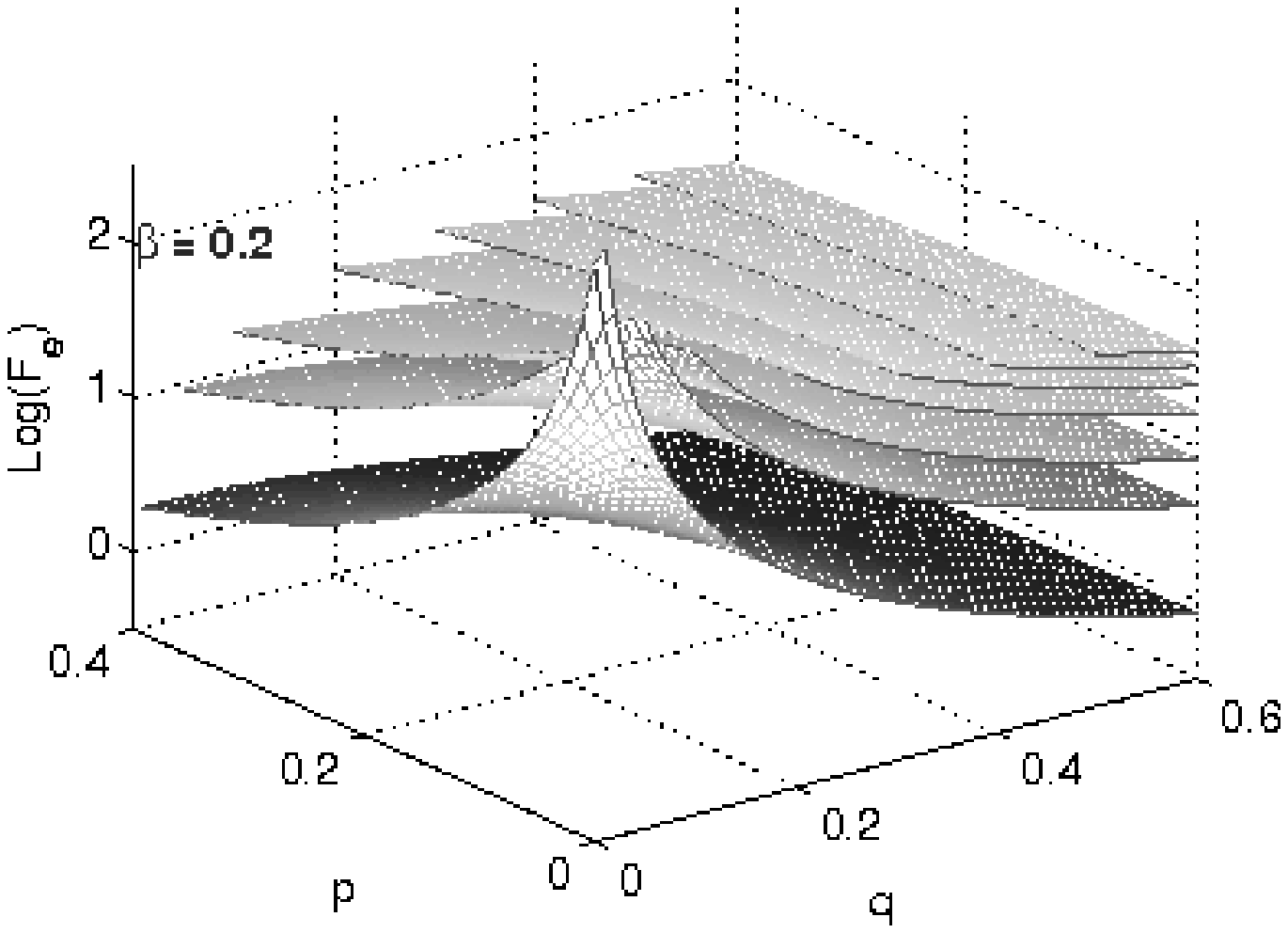}}} }
\caption{$\log_{10} F_e$ for $\beta=.2$.} For further details, see
text.  \label{FigEntro02}
\end{figure}

\newpage

\begin{figure} [htbp]
\centerline{
{\scalebox{0.7}{\includegraphics{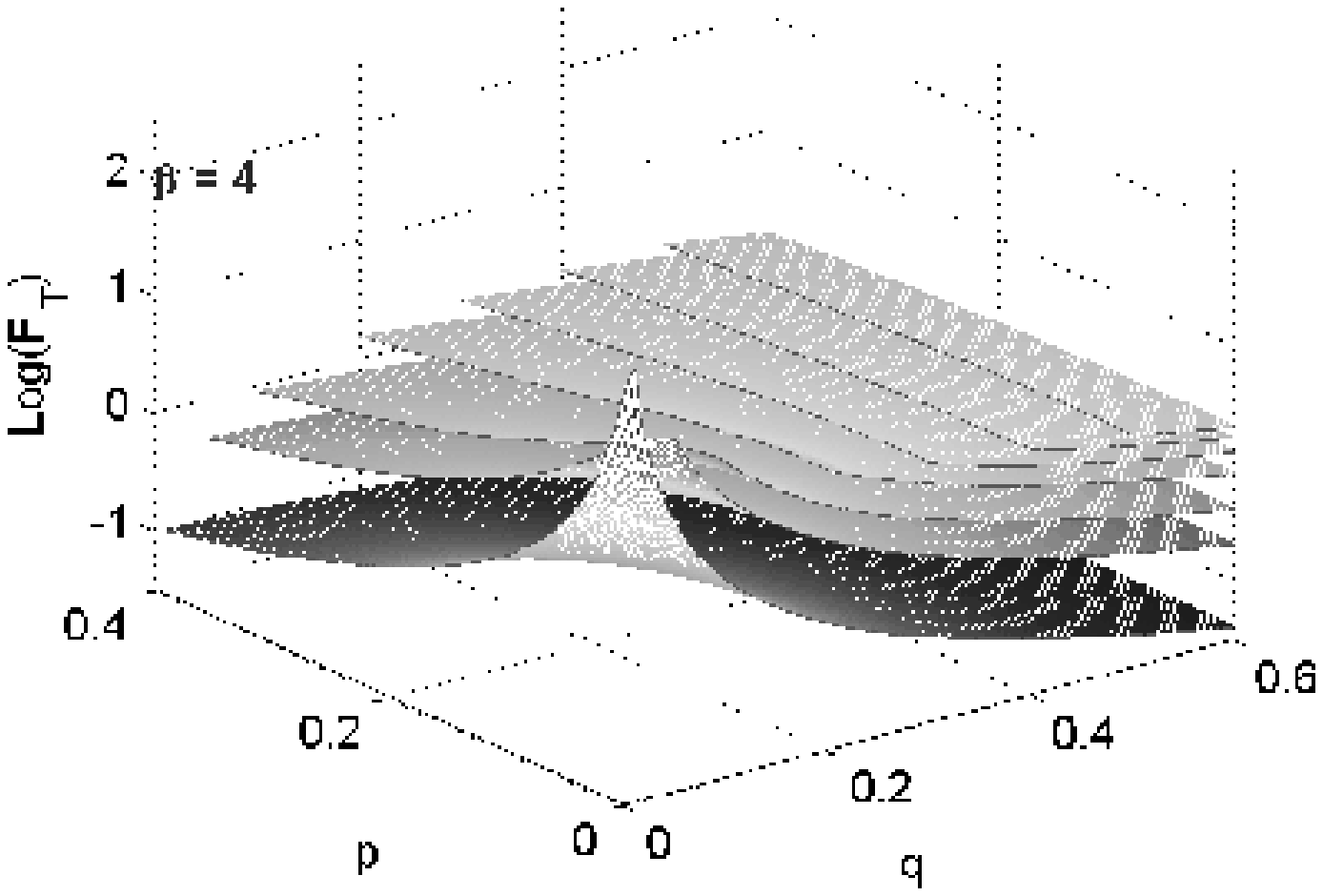}}} }
\caption{$\log_{10} F_T$ for $\beta=4$.} For further details, see
text.  \label{FigTotal40}
\end{figure}

\newpage

\begin{figure} [htbp]
\centerline{
{\scalebox{0.7}{\includegraphics{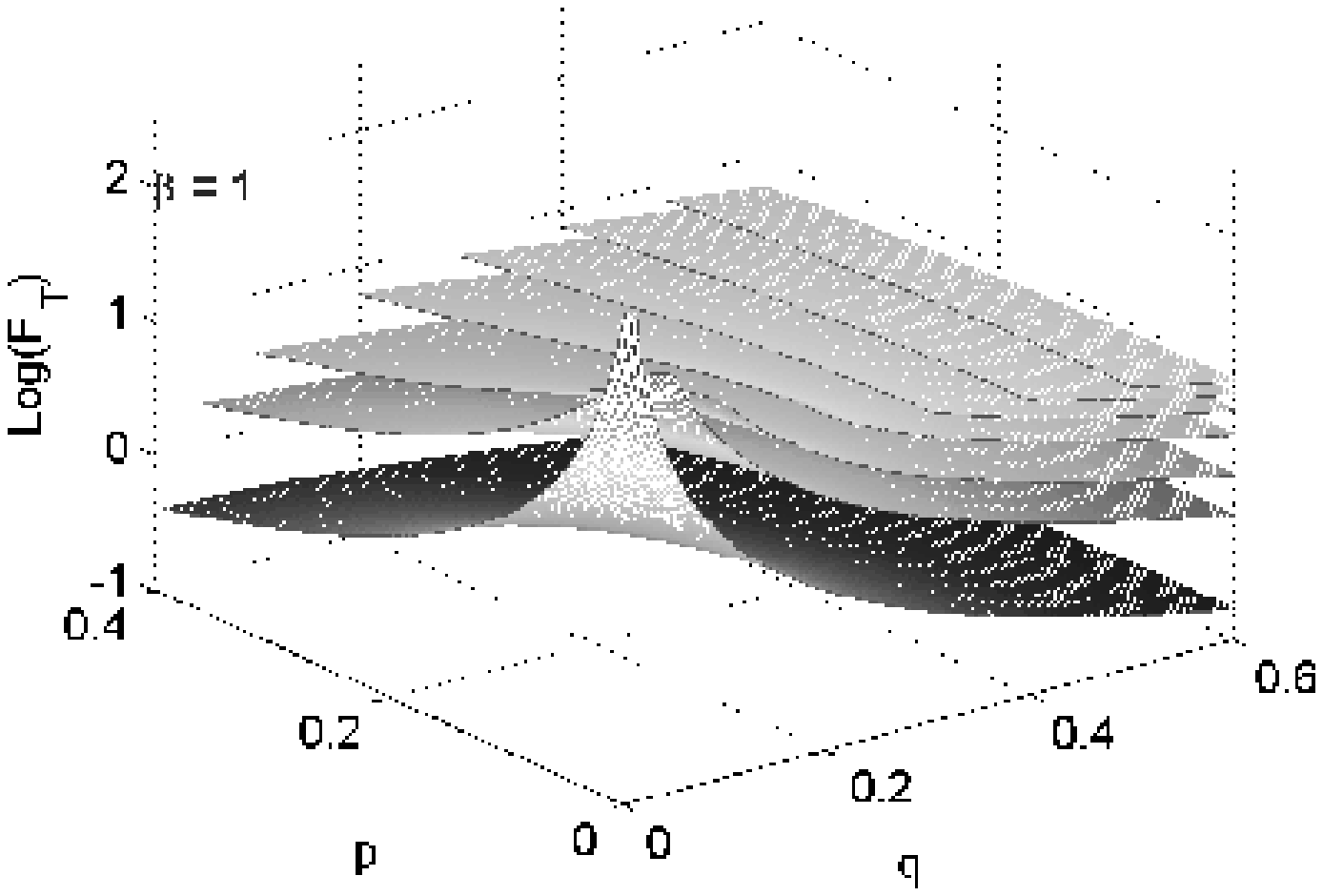}}} }
\caption{$\log_{10} F_T$ for $\beta=1$.}  For further details, see
text. \label{FigTotal10}
\end{figure}

\newpage

\begin{figure} [htbp]
\centerline{
{\scalebox{0.7}{\includegraphics{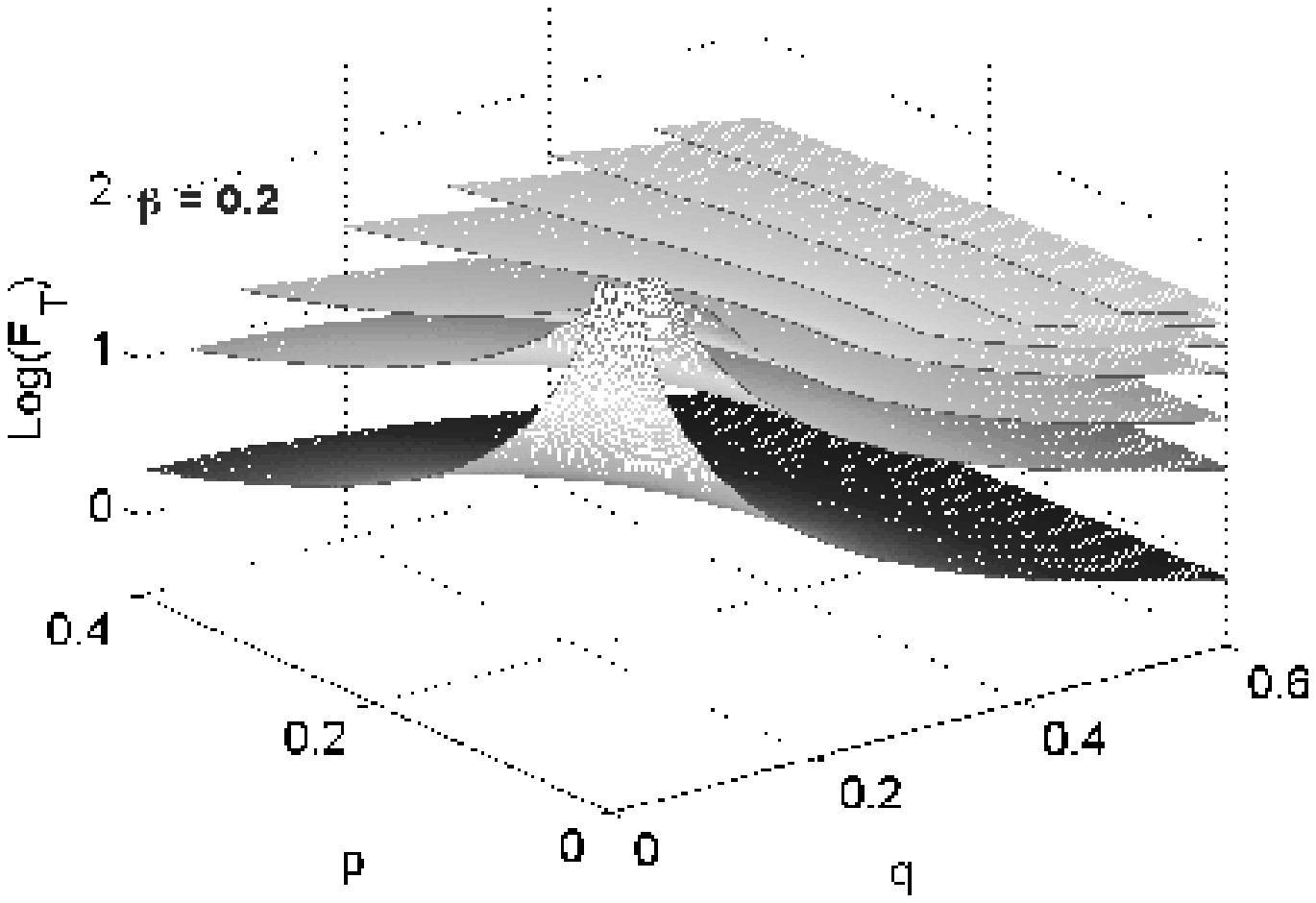}}} }
\caption{$\log_{10} F_T$ for $\beta=.2$.}  For further details,
see text. \label{FigTotal02}
\end{figure}

\newpage

\begin{figure} [htbp]
\centerline{
{\scalebox{0.7}{\includegraphics{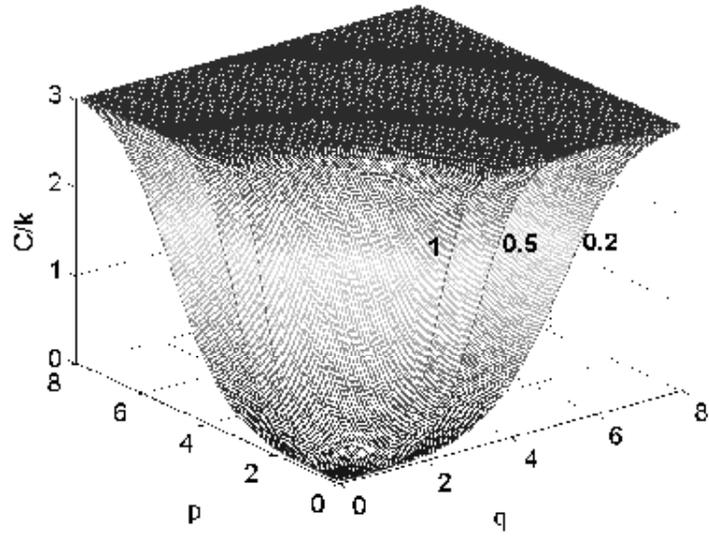}}} } \caption{Specific
heats for, respectively, $\beta: 1, 0.5, 0,2$ (outwards).  For
further details, see text. } \label{FigC}
\end{figure}

\newpage

\begin{figure} [htbp]
\centerline{
{\scalebox{0.7}{\includegraphics{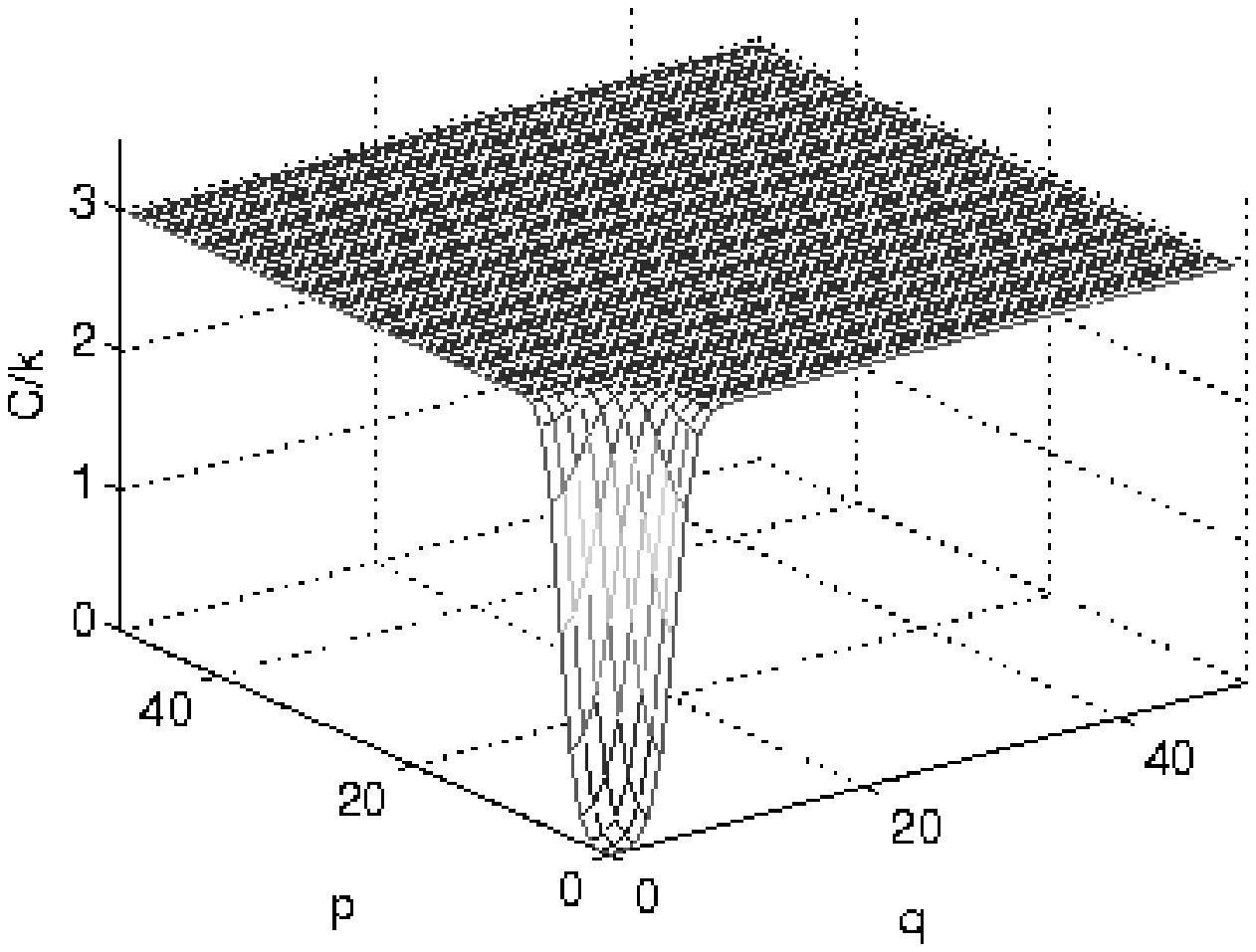}}} }
\caption{Specific heat for $\beta= 0,2$ in an extended
$(q,p)$-range For further details, see text. } \label{FigC2}
\end{figure}

\newpage

\begin{figure} [htbp]
\centerline{
{\scalebox{0.7}{\includegraphics{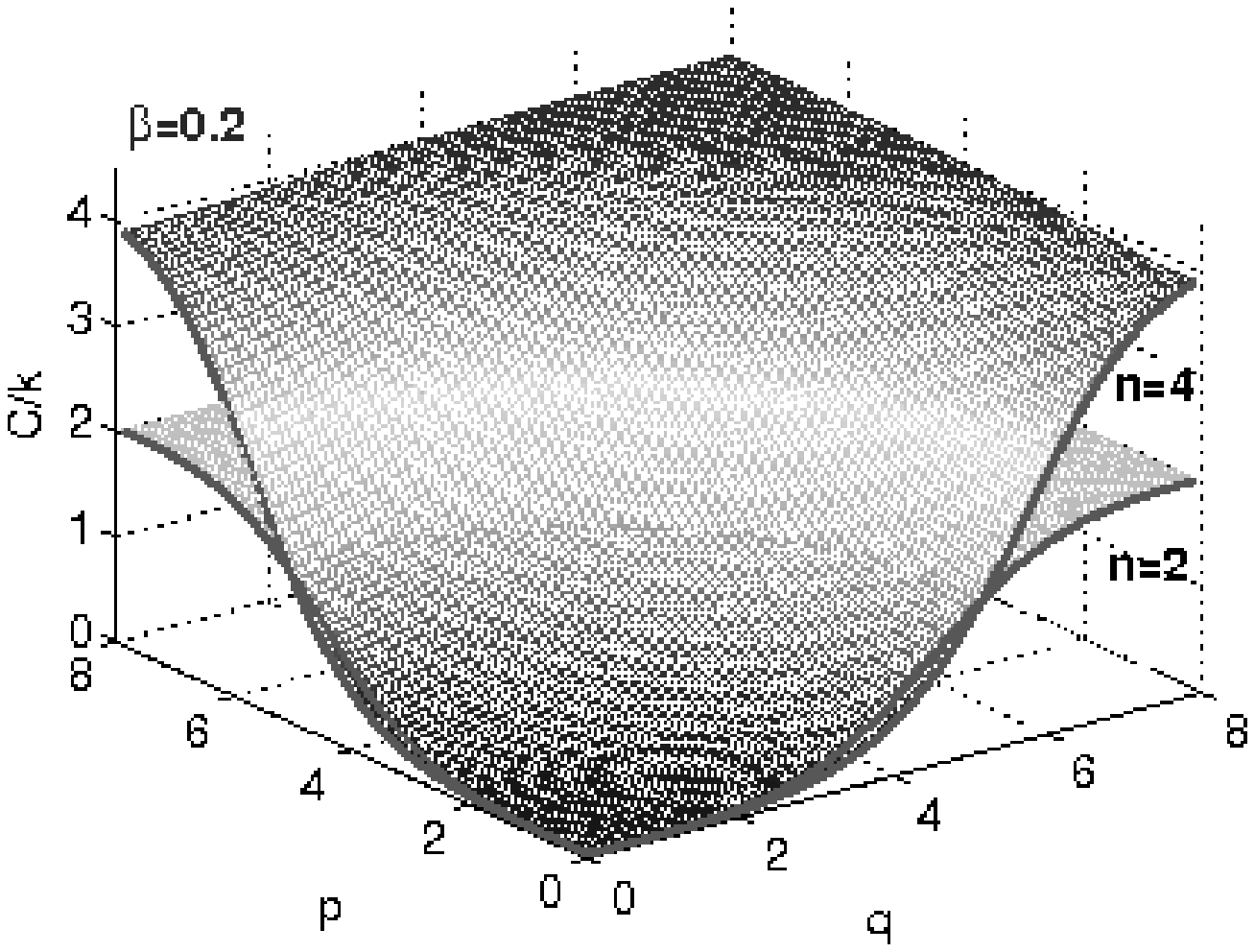}}} }
\caption{Specific heat for $\beta= 0,2$ in an extended
$(q,p)$-range in 2 and 4 dimensions.  For further details, see
text. } \label{FigC2y4}
\end{figure}
\end{document}